\documentclass[twocolumn,pre,reprint,showpacs,floatfix]{revtex4-1}
\usepackage{graphicx,times}

\def\<{\langle}
\def\>{\rangle}
\begin{document}
\title{Improvement of Monte Carlo estimates with covariance-optimized finite-size scaling at fixed phenomenological coupling}
\author{\firstname{Francesco} \surname{Parisen Toldin}}
\affiliation{Max-Planck-Institut f\"ur Physik Komplexer Systeme, N\"othnitzer Strasse 38, D-01187 Dresden}
\email{parisen@pks.mpg.de}

% PACS codes
%05.10.Ln 	Monte Carlo methods
%05.50.+q 	Lattice theory and statistics (Ising, Potts, etc.)
%05.70.Jk 	Critical point phenomena (for quantum critical phenomena in superconductivity, see 74.40.Kb)
%64.60.an 	Finite-size systems

\begin{abstract}
In the finite-size scaling analysis of Monte Carlo data, instead of computing the observables at fixed Hamiltonian parameters, one may choose to keep a renormalization-group invariant quantity, also called phenomenological coupling, fixed at a given value. Within this scheme of finite-size scaling, we exploit the statistical covariance between the observables in a Monte Carlo simulation in order to reduce the statistical errors of the quantities involved in the computation of the critical exponents. This method is general and does not require additional computational time. This approach is demonstrated in the Ising model in two and three dimensions, where large gain factors in CPU time are obtained.
\end{abstract}
\maketitle
The Monte Carlo (MC) method is a powerful numerical technique which allows to study models with a large number of degrees of freedom \cite{Sokal-96,MCbook}. One of the main features of the method is its flexibility: arbitrary statistical systems can be investigated in a regime where other methods (e.g., perturbative calculations) may not be available or efficient. For this reason it is one of the most used tools in critical phenomena \cite{PV-02}.
MC integration allows to compute quantities with, at least in principle, arbitrary precision. The accuracy of the results is proportional to the inverse square root of the computational time, hence even a small improvement in the statistical error bars can represent a significant gain in terms of computational resources. Several strategies have been developed in order to improve the accuracy: one of the most common methods is the use of improved estimators, which have the same expectation value of the desired quantities, but reduced error bars (see, e.g., Ref.~\cite{MCbook}). More recently, it has been proposed to use a covariance analysis in order to compute the optimal weighted average of different estimates of a critical exponent \cite{WJ-09} and to reduce the statistical error of an observable by adding a control variate, whose mean value vanishes \cite{FMM-09}.

In the context of numerical investigations of critical phenomena, finite-size scaling (FSS) techniques play an important role: they allow to extract the critical properties of a model from the data obtained in a region of parameters where the correlation length $\xi$ is of the order of the linear size $L$ of the system. In the FSS analysis of MC data, instead of computing the observables at fixed Hamiltonian parameters, one can choose to fix a renormalization-group (RG) invariant quantity $R(\beta,L)$, also called phenomenological coupling, at a given value $R_f$. In this method, introduced in Ref.~\cite{Hasenbusch-99} and discussed also in Refs.~\cite{HPV-05,HPTPV-07}, one determines for every $L$  an inverse temperature $\beta_f(L)$ such that $R(\beta=\beta_f(L),L)=R_f$. Then all the observables are computed at $\beta=\beta_f(L)$ and the statistical errors can be estimated using standard Jackknife techniques (see, e.g., Ref.~\cite{AMM-book}). It is easy to show that $\beta_f(L\rightarrow\infty)\rightarrow \beta_c$, with $\beta_c$ the inverse critical temperature. It has been observed that, thanks to cross correlations, this method gives in some cases reduced error bars \cite{HPV-05,HPTPV-07}. Error reduction has been reported also in the related FSS method of phenomenological renormalization, where two system sizes are enforced to share a common value of $R=\xi/L$ \cite{BFMMMS-97}.

In this work we study the improvement of the accuracy of MC estimates by considering FSS at fixed phenomenological coupling, to be chosen as a linear combination of a given set of phenomenological couplings. By exploiting the covariance between the observables, we show how to choose the combination which minimizes the statistical errors of the quantities involved in the computation of the critical exponents. We illustrate this method by considering the Ising model in two and three dimensions simulated using, separately, Metropolis and Wolff single-cluster algorithms.

{\bf Finite-size scaling at fixed phenomenological coupling}
A MC simulation consists of a Markov chain whose equilibrium distribution is the Boltzmann-Gibbs measure \cite{Sokal-96}. Any observable is obtained using a statistical estimator, i.e., a random variable whose expectation value is the thermal average of the observable. Since all the observables are calculated from the same configurations generated in the simulation, they are in general correlated with each other.
We denote a random variable with a hat $\widehat{A}$, its expectation value by $E[\widehat{A}]=A$, and its fluctuation around the expectation value by $\widehat{\delta A}\equiv\widehat{A}-A$. In what follows, the dependence of the observables on the size of the system is implicit. Let us consider a generic RG-invariant quantity $R$, which is sampled by an estimator $\widehat{R}$, and that we fix to the value $R_f$:
\begin{equation}
\label{Rfix}
\widehat{R}(\widehat{\beta}_f)=R_f.
\end{equation}
The actual computation of an estimate of $R$, and hence of $\beta_f$, involves an average over the entire MC run, while the statistical error bars resulting from the fluctuations $\widehat{\delta R}$, $\widehat{\delta\beta}_f$ can be estimated using the Jackknife technique. This method consists of dividing the set of MC data into $N_{\rm bin}$ consecutive blocks of equal size, which has to be much larger than the autocorrelation time, and then constructing $N_{\rm bin}$ Jackknife blocks by taking the set of all data except one block. From the data of every single Jackknife block, one computes an estimate of $R$ and of $\beta_f$ by solving Eq.~(\ref{Rfix}), as well as an estimate of any other observable $O$ at the inverse temperature $\beta_f$, as determined from the same Jackknife block. This procedure results in $N_{\rm bin}$ (correlated) estimators $\widehat{O}_f^{\rm (j)}$ of the observable $O$ at fixed $R=R_f$, from which the statistical error bars are computed using the standard Jackknife estimator \cite{AMM-book}:
\begin{equation}
  \label{jackknife}
  \widehat{\sigma^2}(O_f)=\frac{N_{\rm bin}-1}{N_{\rm bin}}\sum_{j=1}^{N_{\rm bin}}\left(\widehat{O}_f^{\rm (j)}-\widehat{O}_f\right)^2,
\end{equation}
where $\widehat{O}_f$ is an estimator of $O$ at fixed $R=R_f$. Note that $\widehat{\sigma^2}(O_f)$ is itself a random variable, whose expectation value is the sought variance. Solving Eq.~(\ref{Rfix}) requires an extrapolation of the MC data sampled at $\beta_{\rm run}$ to $\beta=\widehat{\beta}_f$, which can be done by using reweighting techniques or by computing a Taylor expansion of the observables on $\beta$. This in practice requires $\widehat{\beta}_f\simeq\beta_{run}$, or equivalently $\widehat{R}\simeq R_f$. Employing a first-order Taylor expansion of $\widehat{R}(\widehat{\beta}_f)$ around $\beta_{\rm run}$, $\widehat{\beta}_f$ is obtained as
\begin{equation}
\label{betaf}
\widehat{\beta}_f\simeq\beta_{run} -\frac{\widehat{R}-R_f}{\widehat{R'}},
\end{equation}
where $\widehat{R'}$ is an estimator of the derivative of $R$ with respect to $\beta$. All the other observables are then computed at the inverse temperature $\widehat{\beta}_f$. From an observable $O$ sampled by the estimator $\widehat{O}$ we obtain an observable $O_f$, which is sampled by the random variable $\widehat{O}_f$. Using Eq.~(\ref{betaf}), $\widehat{O}_f$ is given by
\begin{equation}
\label{Of}
\widehat{O}_f\simeq\widehat{O}-\widehat{O'}\frac{\widehat{R}-R_f}{\widehat{R'}},
\end{equation}
where $\widehat{O'}$ is an estimator of the derivative of $O$ with respect to $\beta$ and as before we have used a first-order Taylor expansion of $\widehat{O}$ around $\beta_{\rm run}$. Expanding Eq.~(\ref{Of}) around the expectation value we obtain, to the first order in the fluctuations,
\begin{equation}
\label{Of_fluctuations_full}
\widehat{\delta O}_f\simeq\widehat{\delta O}-\frac{O'}{R'}\widehat{\delta R}-\left(R-R_f\right)\left(\frac{\widehat{\delta O'}}{R'}-\frac{O'}{R'^2}\widehat{\delta R'}\right).
\end{equation}
The second term in Eq.~(\ref{Of_fluctuations_full}) represents the modification in the fluctuation of the observable $O$ which is due to the fact that we have fixed $R$, while the third term is related to the variation in the expectation value due to the extrapolation of the MC data to $\beta=\beta_f$. Since, as mentioned above, $R\simeq R_f$, this last term is a correction, which can be neglected in first approximation.
In the presence of $N$ RG-invariant quantities $\{R_i\}$, $i=1$,\ldots $N$, we can apply the above method by keeping fixed a generic linear combination of $\{R_i\}$, $R=\sum_i \lambda_i R_i$. In this case generalizing Eq.~(\ref{Of_fluctuations_full}) and neglecting its last term, we obtain
\begin{equation}
\label{Of_lambda}
\widehat{\delta O_f}(\{\lambda_i\})=\widehat{\delta O}-\frac{O'\sum_{i=1}^N\lambda_i\widehat{\delta R_i}}{\sum_{i=1}^N\lambda_iR_i'}.
\end{equation}
Now we consider the following problem: find the coefficients $\lambda_i$ which minimize the variance of $\widehat{O_f}(\{\lambda_i\})$. Notice that Eq.~(\ref{Of_lambda}) is invariant under a global rescaling of the coefficients $\lambda_i\rightarrow\mu\lambda_i$, $\mu\neq 0$: this simply reflects the equivalence of fixing $R$ or $\mu R$ . It is convenient to fix the normalization by requiring that
\begin{equation}
\label{constraint}
\sum_{i=1}^N \lambda_i R'_i=O',
\end{equation}
so that we have to minimize the variance of
\begin{equation}
\label{Of_normalized}
\widehat{\delta O}_{f,{\rm norm}}(\{\lambda_i\})=\widehat{\delta O}-\sum_{i=1}^N\lambda_i\widehat{\delta R}_i
\end{equation}
with the constraint of Eq.~(\ref{constraint}). The subscript ${\rm norm}$ in Eq.~(\ref{Of_normalized}) reminds that $\widehat{\delta O}_{f,{\rm norm}}$ is chosen with the normalization given in Eq.~(\ref{constraint}). The optimal $\{\lambda_i\}$ parameters can be found using the method of Lagrangian multipliers. Here, we point out a geometrical interpretation of the problem which allows to find the solution immediately. The space of random variables with null expectation value is a real vector space. A positive-definite scalar product $<,>$ between two vectors $\widehat{\delta A}$ and $\widehat{\delta B}$ can be defined by the covariance ${\rm COV}[\widehat{\delta A}, \widehat{\delta B}]$
\begin{equation}
\label{scalar_product}
<\widehat{\delta A}, \widehat{\delta B}> \equiv {\rm COV}[\widehat{\delta A}, \widehat{\delta B}] = E[\widehat{\delta A}\widehat{\delta B}] - E[\widehat{\delta A}]E[\widehat{\delta B}],
\end{equation}
\begin{figure}
\includegraphics[width=\linewidth,keepaspectratio]{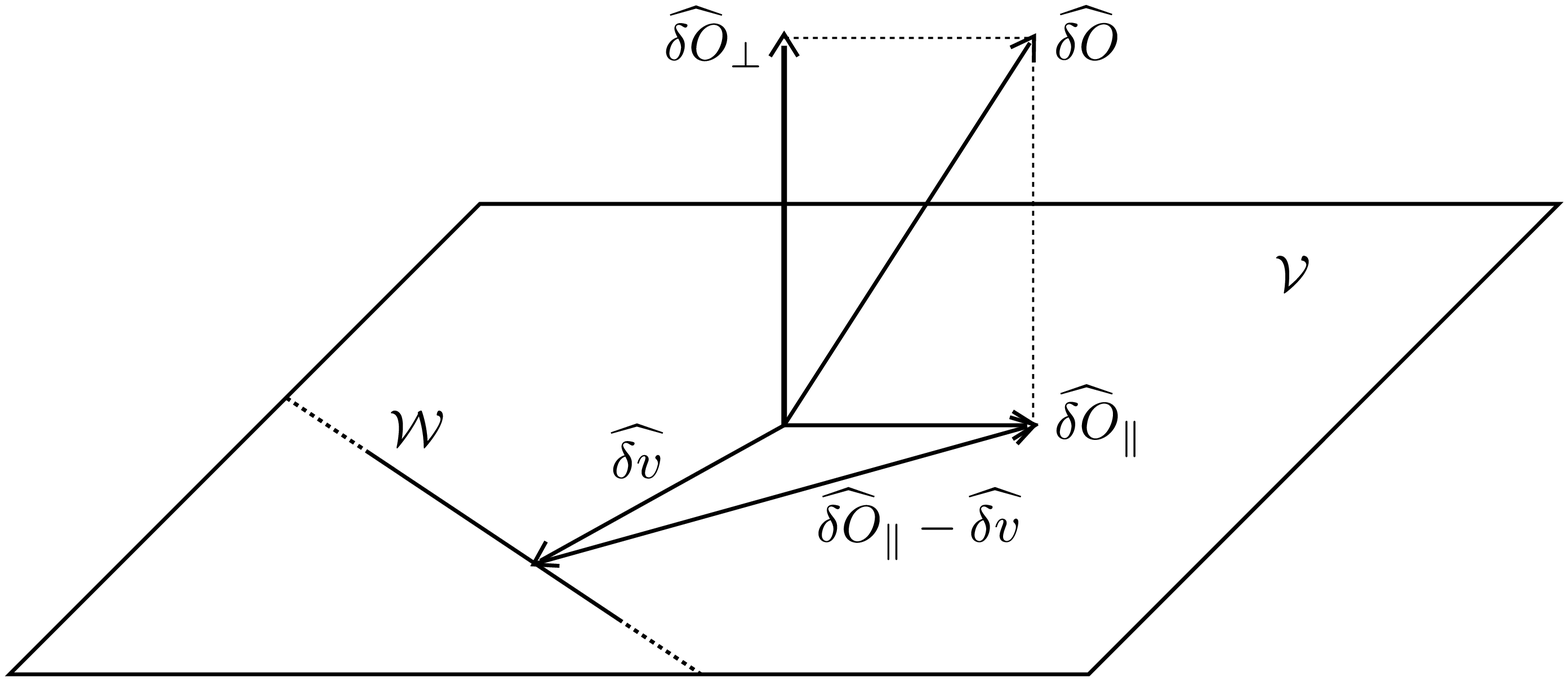}
\caption{Decomposition of the vector $\widehat{\delta O}$ into two components orthogonal $\widehat{\delta O_\perp}$ and parallel $\widehat{\delta O_\parallel}$ to the space \mbox{${\cal V}={\rm Span}<\delta R_1,\ldots,\delta R_N>$}. The affine subspace ${\cal W}$ is given by Eq.~(\ref{constraint}) which represents the normalization on the parameters $\{\lambda_i\}$. The minimum length for $\widehat{\delta O_\parallel}-\widehat{\delta v}$ is obtained when it is orthogonal to ${\cal W}$.}
\label{decomposition}
\end{figure}
The square length of a vector corresponds to the variance of a random variable. Let ${\cal V}={\rm Span}<\widehat{\delta R_1},\ldots,\widehat{\delta R_N}>$ be the $N$-dimensional subspace generated by $\{\widehat{\delta R_i}\}$. According to Eq.~(\ref{Of_normalized}), $\widehat{\delta O}_{f,{\rm norm}}$ is obtained by subtracting to $\widehat{\delta O}$ a vector $\widehat{\delta v}=\sum_i\lambda_i\widehat{\delta R_i} \in {\cal V}$, whose components in the (nonorthogonal) base $\{\widehat{\delta R_i}\}$ are given by the coefficients $\{\lambda_i\}$. By decomposing $\widehat{\delta O}$ into two components orthogonal $\widehat{\delta O_\perp}$ and parallel $\widehat{\delta O_\parallel}$ to $\cal V$, one finds that the variance of $\widehat{\delta O}_{f,{\rm norm}}$ is given by
\begin{equation}
{\rm VAR}[\widehat{\delta O}_{f,{\rm norm}}] = {\rm VAR}[\widehat{\delta O_\perp}] + {\rm VAR}[\widehat{\delta O_\parallel}-\widehat{\delta v}].
\end{equation}
Furthermore, the constraint of Eq.~(\ref{constraint}) represents an affine subspace $\cal W$ of dimension $N-1$, to which $\widehat{\delta v}$ belongs. Minimizing the variance of $\widehat{\delta O}_{f,{\rm norm}}$ consists in finding a vector $\widehat{\delta v}\in {\cal W}$ such that the length of $\widehat{\delta O_\parallel}-\widehat{\delta v}$ is minimum. This situation is illustrated in Fig.~\ref{decomposition}.
\begin{table*}[!t]
\begin{tabular}{l@{}c@{}c@{}c@{}c}
\hline
\hline
     $L$      & $\chi$        & $(dU_4/d\beta)/L^d$   & $(dU_6/d\beta)/L^d$  & $(dR_\xi/d\beta)/L^d$ \\
\hline
Metropolis dynamics&&&&\\
$\bf 8$       & $41.153(99)$  & $-0.06617(99)$  & $-0.1881(30)$  & $0.18353(90)$       \\
imp           & $40.878(29)$  & $-0.06490(17)$  & $-0.18935(44)$ & $0.18074(37)$       \\
gain          & $11.5$        & $35.4$          & $47.6$         & $6.0$               \\

\hline
$\bf 16$      & $140.60(60)$  & $-0.03193(93)$  & $-0.0909(28)$  & $0.09045(73)$       \\
imp           & $138.90(14)$  & $-0.03285(14)$  & $-0.09475(38)$ & $0.08836(34)$       \\
gain          & $18.4$        & $43.6$          & $54.2$         & $4.7$               \\

\hline
$\bf 32$      & $470.8(2.6)$  & $-0.01685(61)$  & $-0.0482(19)$  & $0.04490(42)$       \\
imp           & $469.71(49)$  & $-0.016369(83)$ & $-0.04714(23)$ & $0.04409(20)$       \\
gain          & $27.2$        & $54.9$          & $68.2$         & $4.5$               \\

\hline
$\bf 64$      & $1578.5(5.8)$ & $-0.00812(20)$  & $-0.02334(61)$ & $0.02151(16)$       \\
imp           & $1578.5(1.6)$ & $-0.008121(42)$ & $-0.02336(12)$ & $0.02149(10)$       \\
gain          & $13.3$        & $22.9$          & $27.5$         & $2.6$               \\

\hline
$\bf 128$     & $5284(28)$    & $-0.00433(15)$  & $-0.01251(47)$ & $0.01097(11)$       \\
imp           & $5316.8(6.0)$ & $-0.004078(24)$ & $-0.011736(67)$& $0.010956(62)$      \\
gain          & $22.1$        & $39.8$          & $48.9$         & $2.9$               \\
Wolff dynamics&&&&\\
$\bf 8$       & $41.431(44)$  & $-0.06448(45)$  & $-0.1828(13)$  & $0.18634(60)$       \\
imp           & $40.829(18)$  & $-0.06362(18)$  & $-0.18572(46)$ & $0.18138(38)$       \\
gain          & $5.9$         & $6.3$           & $8.5$          & $2.5$               \\

\hline
$\bf 16$      & $139.70(23)$  & $-0.03294(37)$  & $-0.0942(11)$  & $0.08962(45)$       \\
imp           & $138.000(98)$ & $-0.03237(13)$  & $-0.09353(35)$ & $0.08848(31)$       \\
gain          & $5.4$         & $8.4$           & $9.6$          & $2.2$               \\

\hline
$\bf 32$      & $470.32(84)$  & $-0.01616(21)$  & $-0.04632(63)$ & $0.04359(27)$       \\
imp           & $469.01(35)$  & $-0.016312(77)$ & $-0.04691(21)$ & $0.04362(17)$       \\
gain          & $5.9$         & $7.6$           & $8.7$          & $2.4$               \\

\hline
$\bf 64$      & $1582.8(3.2)$ & $-0.00820(11)$  & $-0.02353(34)$ & $0.02179(15)$       \\
imp           & $1581.4(1.2)$ & $-0.008177(48)$ & $-0.02350(13)$ & $0.02175(10)$       \\
gain          & $6.5$         & $5.7$           & $6.7$          & $2.1$               \\

\hline
$\bf 128$     & $5327(12)$    & $-0.004102(70)$ & $-0.01176(21)$ & $0.010983(89)$      \\
imp           & $5321.2(4.3)$ & $-0.004075(29)$ & $-0.011696(81)$& $0.010883(63)$      \\
gain          & $7.7$         & $5.6$           & $6.5$          & $2.0$               \\
\hline
\hline
\end{tabular}
\caption{Results for the two-dimensional (2D) Ising model with Metropolis and Wolff dynamics, separately. For each lattice size $L$, we compare standard analysis (first line), with the method explained in this work (imp). The approximate gain factor in CPU time is denoted with gain.}
\label{tab:2d}
\end{table*}
The minimum length for $\widehat{\delta O_\parallel}-\widehat{\delta v}$ is then obtained by choosing $\widehat{\delta v}$ such that $\widehat{\delta O_\parallel}-\widehat{\delta v}$ is orthogonal to ${\cal W}$.
With this insight, the optimal coefficients $\{\lambda_i\}$ are calculated as
\begin{equation}
\label{lambda}
\lambda_i = -\frac{{\bf R'^T M^{-1}N}-O'}{\bf R'^TM^{-1}R'}\left({\bf M^{-1}R'}\right)_i+\left({\bf M^{-1}N}\right)_i,
\end{equation}
where the matrix $\bf M$ and the vector ${\bf N}$ are defined as
\begin{eqnarray}
{\bf M}_{ij}\equiv<\widehat{\delta R}_i, \widehat{\delta R}_j>,\\
{\bf N}_i\equiv <\widehat{\delta O}, \widehat{\delta R}_i>,
\label{defN}
\end{eqnarray}
and we have introduced the vector ${\bf R'}_i\equiv R'_i$. The quantities involved in Eq.~(\ref{lambda}) can be estimated from MC data by using standard Jackknife techniques \cite{AMM-book}. It is useful to observe that the covariances which enter in the definitions of ${\bf M}$ and ${\bf N}$ are related to the transition matrix of the Markov chain \cite{Sokal-96}, and hence depend not only on the model, but also on the dynamics. Finally, it is worth noting that in order to correctly obtain the FSS limit, the coefficients $\{\lambda_i\}$ must be the same for every size. They can be, however, chosen differently for each observable $O$ considered.

\begin{table*}
 \begin{tabular}{l@{}c@{}c@{}c@{}c}
\hline
\hline
     $L$      & $\chi$        & $(dU_4/d\beta)/L^d$    & $(dU_6/d\beta)/L^d$  & $(dR_\xi/d\beta)/L^d$ \\
\hline
Metropolis dynamics&&&&\\
$\bf 8$       & $88.19(53)$   & $-0.1112(15)$   & $-0.4758(80)$  & $0.07330(49)$ \\
imp           & $88.90(15)$   & $-0.11412(42)$  & $-0.5341(21)$  & $0.08078(35)$ \\
gain          & $11.7$        & $12.3$          & $14.2$         & $2.0$          \\

\hline
$\bf 16$      & $351.0(3.3)$  & $-0.04140(86)$  & $-0.1813(48)$  & $0.02652(29)$ \\
imp           & $354.91(63)$  & $-0.04269(29)$  & $-0.19932(98)$ & $0.02854(13)$ \\
gain          & $27.9$        & $9.1$           & $24.2$         & $4.9$         \\

\hline
$\bf 32$      & $1382(14)$    & $-0.01574(36)$  & $-0.0697(20)$  & $0.00992(11)$ \\
imp           & $1389.3(2.6)$ & $-0.015837(75)$ & $-0.07383(40)$ & $0.010367(46)$ \\
gain          & $26.9$        & $23$            & $25.8$         & $6.2$          \\

\hline
$\bf 64$       & $5373(46)$   & $-0.00612(12)$  & $-0.02731(70)$ & $0.003686(34)$ \\
imp            & $5435.9(9.8)$& $-0.005953(31)$ & $-0.02760(14)$ & $0.003822(17)$ \\
gain           & $22.2$       & $16.1$          & $25.2$         & $4.2$          \\

\hline
$\bf 128$      & $21537(200)$ & $-0.002220(48)$ & $-0.00981(27)$ & $0.001412(15)$ \\
imp            & $21141(40)$  & $-0.002215(11)$ & $-0.010304(56)$& $0.0014148(61)$ \\
gain           & $25.5$       & $19.7$          & $23.$          & $6.1$           \\
Wolff dynamics&&&&\\
$\bf 8$       & $87.82(28)$   & $-0.1098(11)$   & $-0.4682(55)$  & $0.07244(40)$       \\
imp           & $88.48(11)$   & $-0.11608(37)$  & $-0.5418(20)$  & $0.07995(25)$       \\
gain          & $6.3$         & $8.9$           & $7.7$          & $2.7$               \\

\hline
$\bf 16$      & $350.7(1.2)$  & $-0.04273(45)$  & $-0.1880(23)$  & $0.02704(16)$       \\
imp           & $352.83(47)$  & $-0.04306(17)$  & $-0.20094(90)$ & $0.028609(91)$      \\
gain          & $7.0$         & $7.0$           & $6.7$          & $3.1$               \\

\hline
$\bf 32$      & $1381.2(5.3)$ & $-0.01604(18)$  & $-0.07162(92)$ & $0.009997(63)$      \\
imp           & $1388.4(1.9)$ & $-0.016051(76)$ & $-0.07521(39)$ & $0.010391(36)$      \\
gain          & $7.8$         & $5.6$           & $5.5$          & $3.1$               \\

\hline
$\bf 64$      & $5393(22)$    & $-0.006101(75)$ & $-0.02746(39)$ & $0.003749(27)$      \\
imp           & $5438.6(7.3)$ & $-0.005943(33)$ & $-0.02769(16)$ & $0.003824(15)$      \\
gain          & $9.5$         & $5.3$           & $6.3$          & $3.2$               \\

\hline
$\bf 128$     & $21237(94)$   & $-0.002289(30)$ & $-0.01025(16)$ & $0.001410(10)$      \\
imp           & $21188(29)$   & $-0.002291(14)$ & $-0.010308(64)$& $0.0014145(59)$    \\
gain          & $10.1$        & $4.6$           & $5.8$          & $3.1$               \\
\hline
\hline
\end{tabular}
\caption{Same as in Table \ref{tab:2d} for the 3D Ising model.}
\label{tab:3d}
\end{table*}

{\bf Results}
We have tested this method on the standard Ising model, which is defined on a $d$-dimensional lattice with linear size $L$ by the Hamiltonian
\begin{equation}
\label{ising}
{\cal H} = -J\sum_{\<ij\>}\sigma_i\sigma_j,\qquad \sigma_i=\pm 1,
\end{equation}
where the sum is over the nearest-neighbors sites and one can set $J=1$. We consider here four RG-invariant quantities. Besides the cumulants $U_4$ and $U_6$ defined by
\begin{equation}
\label{U2j}
U_{2j}\equiv \frac{\<M^{2j}\>}{\<M^2\>^j},\qquad M\equiv \sum_i\sigma_i,
\end{equation}
we also consider the ratio of the second-moment correlation length $\xi$ over $L$, $R_\xi\equiv\xi/L$, as well as the ratio of the partition function of a system with antiperiodic boundary conditions (b.c.) on one direction over the partition function of a system with full periodic b.c. $R_Z\equiv Z_a/Z_p$; this quantity can be efficiently sampled using the boundary flip algorithm \cite{Hasenbusch-93}. In Eq.~(\ref{U2j}) the brackets $\<\quad\>$ denote the thermal average. As for the observables, we consider here the susceptibility $\chi\equiv \<M^2\>/L^d$, and the derivatives of RG-invariant quantities $dU_4/d\beta$, $dU_6/d\beta$, $dR_\xi/d\beta$. At the critical point, as well as in the FSS at fixed phenomenological coupling, one has $\chi\propto L^{2-\eta}$ and $dR/d\beta\propto L^{1/\nu}$, $R=U_4$,$U_6$,$R_\xi$, where $\eta$ and $\nu$ are critical exponents. Thus these observables can be used to extract the value of the critical exponents. As mentioned after Eq.~(\ref{defN}), the coefficients $\{\lambda_i\}$ can be optimized separately for each observable. However, for a given observable they have to be the same for each lattice size. Of course, the coefficients determined from Eq.~(\ref{lambda}) will be in general different for each lattice. Since in a typical simulation most of the computational time is spent on the largest lattice, a possible strategy for choosing the coefficients $\{\lambda_i\}$ consists in applying Eq.~(\ref{lambda}) to the largest available lattice, and then to use the same coefficients for the smaller lattices. We have found that the gain in the CPU time in most of the cases does not vary significantly with the lattice size, despite the fact that the chosen $\{\lambda_i\}$ are not the optimal ones for the smaller lattices.
We have simulated the Ising model for lattice sizes $L=8$--$128$ in two and three dimensions. As mentioned above, the statistical error bars and hence the gain in the CPU time resulting from the present method depend not only on the model, but also on the dynamics used. To illustrate this aspect, we have considered data sampled using, separately, two different simulation algorithms: Metropolis and Wolff single cluster.
Simulations in $d=2$ have been carried out at $\beta=0.4406867935$. The analysis at fixed phenomenological coupling has been done using the critical-point values $U_4=1.167923(5)$, $U_6=1.455649(7)$, $R_\xi=0.9050488292(4)$ \cite{SS-00}, and $R_Z\simeq 0.3728848808$ \cite{CH-96}. The results are reported in Table~\ref{tab:2d}. We compare the observables as determined with a standard analysis and with the present method (imp), with the coefficients determined, for each observable, from the data at $L=128$.
Notice that the mean values are not expected to coincide in the two analyses, although they can be very close. The gain factor in the CPU time is given by the square of the ratio of the error bars.
Inspecting Table~\ref{tab:2d} we observe large gains in the CPU time, especially for the observables $\chi$, $dU_4/d\beta$, $dU_6/d\beta$ sampled with the Metropolis algorithm, where in most of the cases the gain lies between $20$ and $50$; for the same observables, gains roughly between $6$ and $9$ are obtained in the case of Wolff dynamics. The gain in the CPU time obtained for $dR_\xi/d\beta$ is roughly $\sim 2$--$3$ for both algorithms. Simulations in $d=3$ have been carried out at $\beta=0.2216544$. FSS analysis at fixed phenomenological coupling has been done using recent determinations of the critical-point values $U_4=1.6036(1)$, $U_6=3.1053(5)$, $R_\xi=0.6431(1)$, $R_Z=0.5425(1)$ \cite{Hasenbusch-10}. The results are reported in Table~\ref{tab:3d}. The gains in the CPU time for $\chi$, $dU_4/d\beta$, $dU_6/d\beta$ sampled with the Metropolis dynamics are in most of the cases between $20$ and $30$. For the other observables and for the Wolff dynamics the gains are similar to the bidimensional case.
The present method represents also a significant improvement over a simpler approach of fixing one of the four RG-invariant quantities: for instance by fixing $R_\xi$ the CPU gain factor for $\chi$ is $\lesssim 4$ in $d=2$, $\lesssim 9$ in $d=3$ with Metropolis dynamics, and $\lesssim 6$ in $d=3$ with Wolff dynamics.

To summarize, we have presented a FSS method which allows for a substantial reduction of the statistical error bars without additional computational time. Application of the method to the Ising model shows that the CPU gain factor does not depend much on the lattice size, and it is more pronounced for a local Metropolis update algorithm. Other possible interesting applications of this method should be found in ``improved models'' \cite{PV-02}, where leading scaling corrections are suppressed, and in models with quenched disorder \cite{KR-04}, where cluster algorithms are available only in special cases and significant reduction of error bars with FSS at fixed $R_\xi$ has been reported \cite{HPTPV-07}.
\begin{acknowledgments}
The author is grateful to Martin Hasenbusch for useful discussions and communications and to Roderich Moessner and Ettore Vicari for useful comments on the manuscript.
The author acknowledges support from the Max-Planck-Institut f\"ur Metallforschung (Stuttgart, Germany), where part of the work has been done.
\end{acknowledgments}

\end{document}